\documentclass[preprint]{aastex}
\usepackage{graphicx}
\newcommand {\be} {\begin{equation}}
\newcommand {\ee} {\end{equation}}

\begin{document}
 
\title{Inferring the Magnetic Fields of Magnetars from their X-ray Spectra}

\author{Tolga G\"uver$^{1,2}$, Feryal \"Ozel$^{1}$ and Maxim Lyutikov$^{3}$}

\affil{$^{1}$University of Arizona, Department of Physics,  1118 E. 4th
St., Tucson, AZ 85721} 
\affil{$^{2}$Istanbul University, Astronomy \& Space 
Sciences Department, Beyaz\i t, Istanbul, 34119} 
\affil{$^{3}$Department of Physics, Purdue University, 525 Northwestern 
Avenue, West Lafayette, IN 47907-2036}

\begin{abstract}

We present the first self-consistent theoretical models of magnetar
spectra that take into account the combined effects of the stellar
atmosphere and its magnetosphere. We find that the proton cyclotron
lines that are already weakened by atmospheric effects become
indistinguishable from the continuum for moderate scattering optical
depths in the magnetosphere. Furthermore, the hard excess becomes more
pronounced due to resonant scattering and the resulting spectra
closely resemble the observed magnetar spectra. We argue that while
the absence of proton cyclotron lines in the observed spectra are
inconclusive about the surface field strengths of magnetars, the
continuum carries nearly unique signatures of the field strength and
can thus be used to infer this quantity. We fit our theoretical
spectra with a phenomenological two-blackbody model and compare our
findings to source spectra with existing two-blackbody fits. The field
strengths that we infer spectroscopically are in remarkable agreement
with the values inferred from the period derivative of the sources
assuming a dipole spindown. These detailed predictions of line
energies and equivalent widths may provide an optimal opportunity for
measuring directly the surface field strengths of magnetars with
future X-ray telescopes such as {\em Constellation}-X.

\end{abstract}
 
\section{Introduction}
Anomalous X-ray Pulsars (AXPs) and Soft Gamma-ray Repeaters (SGRs) are
two similar classes of neutron stars thought to be powered by
ultrastrong magnetic fields \citep{d1992}. Initially identified by
their pulsed soft X-ray \citep{f1981} and $\gamma$-ray emissions
\citep{m1981}, respectively, they are by now observed in multiple
wavebands, exhibiting a variety of interesting and complex phenomena
(see, e.g., \citealp{w2004,w2006,c2006}). However,
their persistent thermal-like X-ray spectra (with kT $\sim$ 0.3-0.6
keV) and their X-ray luminosities ($L_{X} \sim 10^{34-36}$ ergs/s)
that significantly exceed their spin-down luminosities remain as a
defining characteristic (e.g., \citealt{m1995}) and has been the
subject of numerous studies.

Empirical models of the observed X-ray spectra of AXP \& SGRs have
typically required, in addition to a blackbody component, a power-law
component, with photon indices ranging between 2.0 and 4.5, or a
second blackbody (see, e.g., \citealt{w2004}). Although such
modelling is adequate for descriptive purposes, more detailed and
physically motivated calculations have recently been performed,
focusing on the effects of either neutron star atmospheres or their
magnetospheres on the X-ray spectra.

Radiative equilibrium models of highly magnetized neutron star
atmospheres have been developed recently by numerous authors focusing
on different physical processes. Full angle- and
polarization-dependent scattering has been incorporated by
\citet{o2001} and \citet{l2003}.  The effects of vacuum polarization
resonance were treated completely, including the resonant coupling of
modes, by \citet{o2001} and \citet{l2003}, and approximately by
\citet{ho2003} through a probabilistic mode conversion approach. Ion
cyclotron lines were calculated by \citet{z2001}, \citet{ho2001}, and
\citet{o2003}.  Finally, the effects of the two-dimensional structure
of the magnetic field were considered by \citet{l2003} and partial
ionization in the atmosphere by \citet{ho2003}. Most models predict
significant deviations from a Planckian spectrum with hard excess that
depends on the magnetic field strength as well as absorption features
at the ion cylotron frequency that are somewhat suppressed by the
vacuum polarization resonance.
 
The surface emission from a neutron star is processed by its
magnetosphere before it reaches a distant observer. \citet{g1969}
showed that there is a minimum plasma density given by
$n=7\times10^{-2}\it{B}/\it{P}$ $\rm{cm}^{-3}$ in the magnetosphere of
a neutron star (where $\it{B}$ is the magnetic field strength in Gauss
and $\it{P}$ is the stellar spin period in seconds). If the
magnetospheric plasma density is equal to this minimum Goldreich \&
Julian density, \citet{m1982} showed that scattering of the surface
emission by such a plasma is not important in the X-ray
regime. However, in the case of magnetars, \citet{t2002} showed that
large-scale currents flowing in the magnetosphere result in much
larger particle densities such that resonant cyclotron scattering can
be an effective mechanism to modify the emitted spectrum from the
surface of a neutron star. This has been proposed as an alternative to
the surface models to explain the power-law tail of X-ray spectra of
AXPs and SGRs. Inspired by this idea, \citet{l2006} calculated the
effects of resonant cyclotron scattering on a photon emitted by a
Planckian source in the magnetosphere of a neutron star. They showed
that the initial spectrum can be shifted to higher energies, which
gives rise to a high-energy tail.

In the magnetar context, the combined signatures of the atmosphere and
the magnetosphere on the spectrum have not been studied to date. In
this paper we calculate the emitted spectra from the surface of a
highly magnetized neutron star by incorporating all the physical
processes that take place in its atmosphere as well as the resonant
cyclotron scattering in its magnetosphere. We study the effects of
scattering, both on the continuum of the spectrum and the proton
cyclotron lines. We use our results to argue that the absence of the
proton cyclotron line is not a good indicator of the stellar field
strength while the shape of the continuum can be used to infer a
combination of the effective temperature and the surface magnetic
field strength. 

\section{Models}

We follow the methods discussed in \citep{o2001,o2003} and
\citet{l2006} to calculate the emission from a neutron star atmosphere
in radiative equilibrium and the resonant cyclotron scattering of
photons in the magnetosphere, respectively.

The neutron star atmosphere models are calculated for fully-ionized H
plasmas, taking into account the effects of vacuum polarization and
ion cyclotron lines. Here, we also incorporate the higher harmonics of
the ion cyclotron lines, following \citet{g2006}. As in
the previous papers, we construct radiative equilibrium atmospheres
using a modified Lucy-Uns\"old algorithm. The surface emission
spectrum is completely defined by the effective temperature
$T_{\rm{eff}}$ of the atmosphere, the surface magnetic field strength
$B$, and the gravitational acceleration $g$ on the stellar
surface. Throughout this paper, we set g=1.9$\times10^{14}$
cm$\,\rm{s}^{-2}$.

We calculate the effect of the resonant scattering on the spectrum
using the Green's function approach described in \citet{l2006}. 
We assume that the magnetic field in the magnetosphere is
spherically symmetric, following a $1/r^{3}$ dependence. We solve the
radiative transfer equation using a Schwarzschild-Schuster two-stream
approximation. The emerging spectrum depends on two parameters: the
resonant scattering optical depth
\be
\tau \equiv \frac{\pi^{2}e^{2}nr}{3mc\omega_{B}} 
\ee
and $\beta$, the thermal electron velocity in units of the speed of
light. Here, $n$ is the plasma density, $r$ is the radius of the
cyclotron scattering region for a given photon energy, $\omega_{B}
\equiv eB/mc$ is the cyclotron frequency, e and m are the electron
charge and mass, respectively.

\section{Results}

In order to investigate the effects of scattering on the surface
emission spectrum, we have computed model spectra for surface
temperatures 0.1-0.5 keV, and for magnetic field strengths
covering a range from $10^{13}$ to $10^{15}$ G. We have then processed
the emission spectrum through the magnetosphere using the resonant
cyclotron scattering model, for optical depth values $\tau$ = 1 - 10
and electron velocity values $\beta$ = 0.1 - 0.5.

\begin{figure*}
\centering
   \includegraphics[scale=0.3]{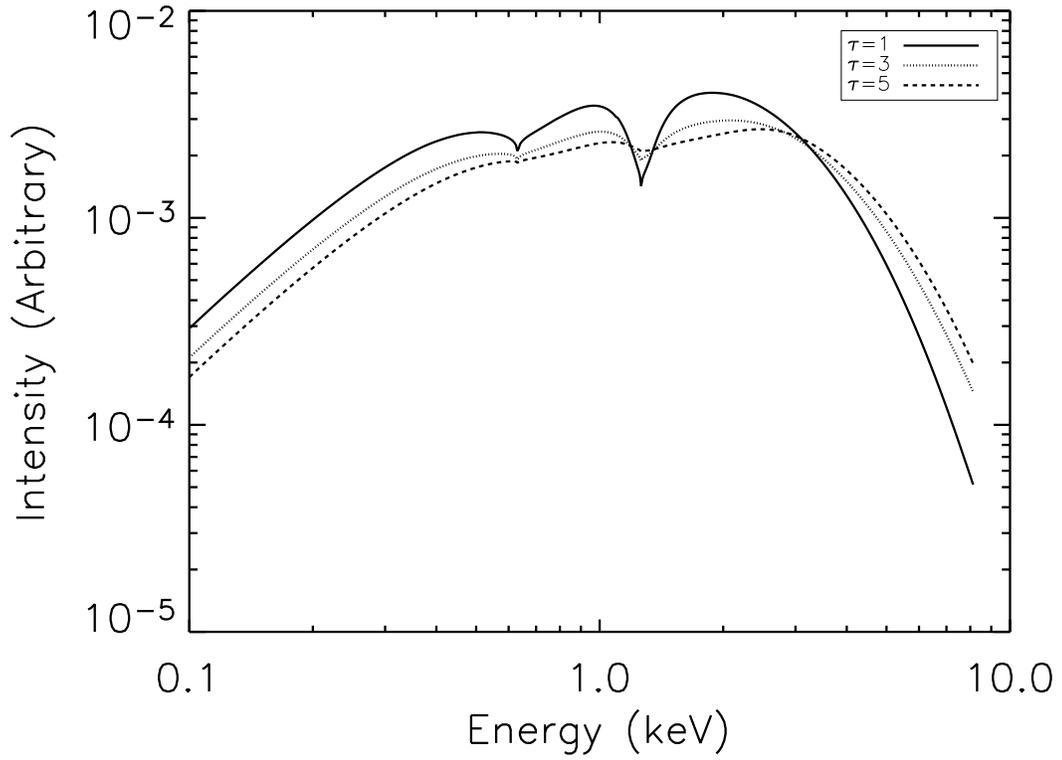}
   \caption{The spectra emerging from a magnetar with a surface
magnetic field strength of $10^{14}$~G, a surface temperature of
0.3~keV, and different scattering optical depths in the
magnetosphere. The equivalent width of the proton cyclotron line
decreases and the spectrum at high energies hardens with increasing
optical depth.}

\label{cc1}
\end{figure*}

To illustrate the effects of resonant cyclotron scattering on both the
continuum spectrum and the proton cyclotron lines, we plot, in Figure
\ref{cc1}, the spectrum emerging from a neutron star with 0.3 keV
surface temperature, $10^{14}$ G surface magnetic field strength, and
increasing scattering optical depth in the magnetosphere.  At photon
energies above the peak of the spectrum, the non-Planckian tails
already present in the surface spectra become harder. This is caused
by the energy gain experienced by the photons at each resonant
scattering of the electrons.  The deep proton cyclotron absorption
features present in the surface spectra are smoothed out as a result
of the effect of resonant cyclotron scattering, as also pointed out by
\citet{l2006}. This is in addition to the fact that
vacuum polarization resonance in the atmosphere already reduces the
equivalent widths of the proton cyclotron features \cite{o2001}. 
Note that in Figure \ref{cc1}, we have chosen a surface
emission spectrum with particularly strong harmonics of the proton
cyclotron line to demonstrate the effects discussed here.

\begin{figure*}
\centering
   \includegraphics[scale=0.3]{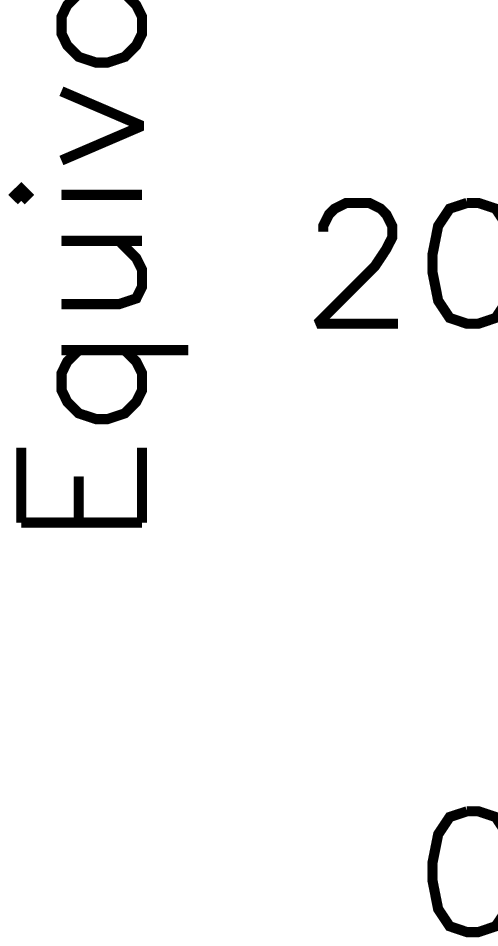}
   \caption{The change in the equivalent width of the fundamental
   proton cyclotron line as a function of optical depth $\tau$, for
   different values of the electron velocity $\beta$. + sign denotes
   the equivalent width of the line when no scattering taken into
   account. The atmospheric calculation is for a magnetic field
   strength of $4\times10^{14}$ Gauss and a surface temperature of 0.3
   keV.}
\label{eq1}
\end{figure*}

In Figure \ref{eq1} and Table \ref{t_eq}, we show quantitatively the
reduction of the equivalent width of the fundamental proton cyclotron
line caused by magnetospheric scattering. The atmosphere model
calculation in Figure \ref{eq1} is for a neutron star with a surface
magnetic field of $10^{14}$ G and a surface temperature of 0.3~keV.
Results for different magnetic field strengths and the harmonics of
the cyclotron lines are presented in Table \ref{t_eq}. As can be seen
in Figure \ref{eq1}, the equivalent widths depend very weakly on the
electron velocity $\beta$, but rapidly decrease with scattering
optical depth $\tau$. Even a scattering optical depth of unity is
enough to reduce the equivalent width of the lines almost by half. For
the model parameters used here, this optical depth corresponds to a
particle density of $\sim 3 \times 10^{16}$ $\rm{cm}^{-3}$, which
is $\sim10^{4}$ times larger than the Goldreich-Julian density for a
$10^{14}$~G magnetar with a period of 6s.

\begin{deluxetable}{ccccc}
\tablecolumns{5}
\tablewidth{280pt}
\tablecaption{Cyclotron line equivalent widths\tablenotemark{a}}
\tablehead{
Magnetic Field  & Line Energy   & \multicolumn{3}{c}{Equivalent Width (eV)}	 \\
($10^{14}$~G)&    (keV)      & $\tau = 1$ & $\tau = 3$      & $\tau = 5$ }
\startdata
0.1		&	0.063	&    0.87    & 0.44	&	-\tablenotemark{b} \\
0.1		&	0.126	&    0.80    & 0.41	&	-\tablenotemark{b} \\
1		&	0.635	&    27.96   & 14.61   & 6.7 \\
1	        &       1.263	&    19.78   & 9.54   & 3.9 \\
4		&       2.529	&    35.05   & 12.02    & -\tablenotemark{b} \\	
4		&	5.051	&    130.29  & 44.07   & 14.6 \\
8		&	5.050	&    129.68  & 27.56   & -\tablenotemark{b} \\
10		&	6.295   &    122.77  & -\tablenotemark{b}	& -\tablenotemark{b} \\	
\enddata
\tablenotetext{a}{The electron velocity is $\beta=0.3$.}
\tablenotetext{b}{The lines are indistinguishable from the continuum spectra.}
\label{t_eq}
\end{deluxetable}

Both vacuum polarization and resonant cyclotron scattering in the
magnetosphere have similar effects on the X-ray spectrum of a
magnetar. These effects give rise to both reduced widths of cyclotron
lines and to non-planckian shapes in the high energy part of the
spectrum ($\sim$2.0-10.0~keV). Usually this component is modeled
phenomenologically by adding a second blackbody or a power law to a
thermal model. In Table \ref{t_con}, we show the results of fitting
our calculated spectra with a phenomenological two-blackbody
model. Typically, the temperature of the soft blackbody corresponds to
the surface temperature of the neutron star, whereas the second
blackbody is used to fit the high-energy tail of the spectrum.

\begin{deluxetable}{cccccccc}
\tablecolumns{8}
\tablewidth{360pt}
\tabletypesize{\small}
\tablecaption{Phenomenological two-blackbody description of magnetar
models\tablenotemark{a}}
\tablehead{
Magnetic Field & Surface Temp.& \multicolumn{2}{c}{$\tau = 1$} & \multicolumn{2}{c}{$\tau = 3$}  
& \multicolumn{2}{c}{$\tau = 5$} \\
($10^{14}$ G)&    (keV)       & $kT_{1}$ & $kT_{2}$ & $kT_{1}$ & $kT_{2}$ & $kT_{1}$ & $kT_{2}$ }
\startdata
1		& 0.1           & 0.11	& 0.36	& 0.12	& 0.42	& 0.13	& 0.60	\\
2		& 0.1		& 0.08	& 0.14	& 0.10	& 0.19	& 0.10	& 0.20	\\
4		& 0.1		& 0.09	& 0.18	& 0.09	& 0.21	& 0.10	& 0.23	\\
6		& 0.1		& 0.09	& 0.20	& 0.10	& 0.22	& 0.10	& 0.25	\\
8		& 0.1		& 0.09	& 0.20	& 0.10	& 0.23	& 0.10	& 0.25	\\
10		& 0.1           & 0.09	& 0.20	& 0.09	& 0.22	& 0.10	& 0.24	\\   
1		& 0.2		& 0.17	& 0.51	& 0.18	& 0.61	& 0.19	& 0.69	\\
2		& 0.2		& 0.22	& 0.66	& 0.23	& 0.73	& 0.25	& 1.07	\\
4		& 0.2		& 0.11	& 0.27	& 0.14	& 0.32	& 0.16	& 0.35	\\
6		& 0.2		& 0.15	& 0.32	& 0.17	& 0.37	& 0.19	& 0.41	\\
8		& 0.2		& 0.17	& 0.35	& 0.18	& 0.40	& 0.19	& 0.44	\\
10		& 0.2           & 0.17	& 0.36	& 0.18	& 0.41	& 0.19	& 0.45	\\
1		& 0.3           & 0.22	& 0.62	& 0.24	& 0.75	& 0.26	& 0.83  \\ 
2		& 0.3		& 0.31	& 0.83	& 0.32	& 0.90	& 0.32	& 0.99	\\ 
4		& 0.3		& 0.35	& --\tablenotemark{b}	& 0.38	&--\tablenotemark{b}	
& 0.43	&--\tablenotemark{b}	\\
6		& 0.3		& 0.16	& 0.38	& 0.21	& 0.44	& 0.24	& 0.49	\\
8		& 0.3		& 0.19	& 0.41	& 0.23	& 0.48	& 0.25	& 0.53	\\
10		& 0.3           & 0.22	& 0.44	& 0.24	& 0.51	& 0.26	& 0.56	\\
1		& 0.4		& 0.25 	& 0.70 	& 0.25 	& 0.80	& 0.27 	& 0.88 	\\
2		& 0.4		& 0.33	& 0.86	& 0.35	& 0.92	& 0.38	& 0.99	\\
4		& 0.4		& 0.49	&--\tablenotemark{b}	& 0.51	&--\tablenotemark{b}	
& 0.54	&--\tablenotemark{b}	\\ 
6		& 0.4		& 0.46	&--\tablenotemark{b}	& 0.51	&--\tablenotemark{b}	
& 0.55	&--\tablenotemark{b}	\\
8		& 0.4		& 0.46	&--\tablenotemark{b}	& 0.27	& 0.57	& 0.31	& 0.61	\\ 
10		& 0.4           & 0.20	& 0.48	& 0.28	& 0.57	& 0.31	& 0.63	\\   
1		& 0.5           & 0.26	& 0.75	& 0.31	& 0.87	& 0.34	& 0.96	\\
2		& 0.5		& 0.44	& 1.01	& 0.44	& 1.07	& 0.51	& 1.23	\\
4		& 0.5		& 0.60	& 1.32	& 0.60	& 1.47	& 0.63	& 1.60	\\
6		& 0.5		& 0.53	&--\tablenotemark{b}	& 0.58	&--\tablenotemark{b}	
& 0.63	&--\tablenotemark{b}	\\
8		& 0.5		& 0.54	&--\tablenotemark{b}	& 0.59	&--\tablenotemark{b}	
& 0.65	&--\tablenotemark{b}	\\
10		& 0.5           & 0.56	&--\tablenotemark{b}	& 0.30	& 0.63	& 0.35	& 0.70	\\
\enddata
\tablenotetext{a}{The electron velocity is $\beta=0.3$.}
\tablenotetext{b}{A second blackbody is not required.}
\label{t_con}
\end{deluxetable}

\section{Discussion}

The X-ray spectra of magnetars carry signatures of their magnetic
field strengths, which determine both the shape of their continuum and
the energies and equivalent widths of cyclotron lines. While the
detection of cyclotron lines in AXPs and SGRs would yield the most
direct measurement of their surface magnetic field strengths, numerous
recent searches with Chandra and XMM-Newton have typically resulted
only in upper limits that are given in Table \ref{uplim}. The only exception has
been the absorption line feature reported for SGR 0526$-$66 by \citet{i2003}.

Proton cyclotron lines that are more prominent than the current
detection limits are predicted by all models of the atmospheric
emission from magnetars despite the presence of processes within the
stellar atmospheres that reduce their strengths (see,
e.g., \citealt{z2001,ho2003,o2003}).  This has been used
to argue that the surface field strengths of these sources are such
that the cyclotron lines are outside of the observed energy range and
may be different than the field strengths inferred from the dipole
spin-down formula (e.g., \citealt{t2005}).

The theoretical predictions, however, are different when we consider
the effects of the magnetosphere. As we showed in \S3, the proton
cyclotron lines are smeared and often become indistinguishable from
the continuum even for moderate scattering optical depths. This is
because the scattering of the surface photons in the resonant layers
in the magnetosphere broadens any spectral feature due to the random
motions of the resonant electrons.  It is, therefore, not surprising
that searches for proton cyclotron lines typically yield no
detections. This also shows that the absence of cyclotron lines is not
a good indicator of the magnetic field strength of the neutron star.

\begin{figure*}
\centering
   \includegraphics[scale=1.0]{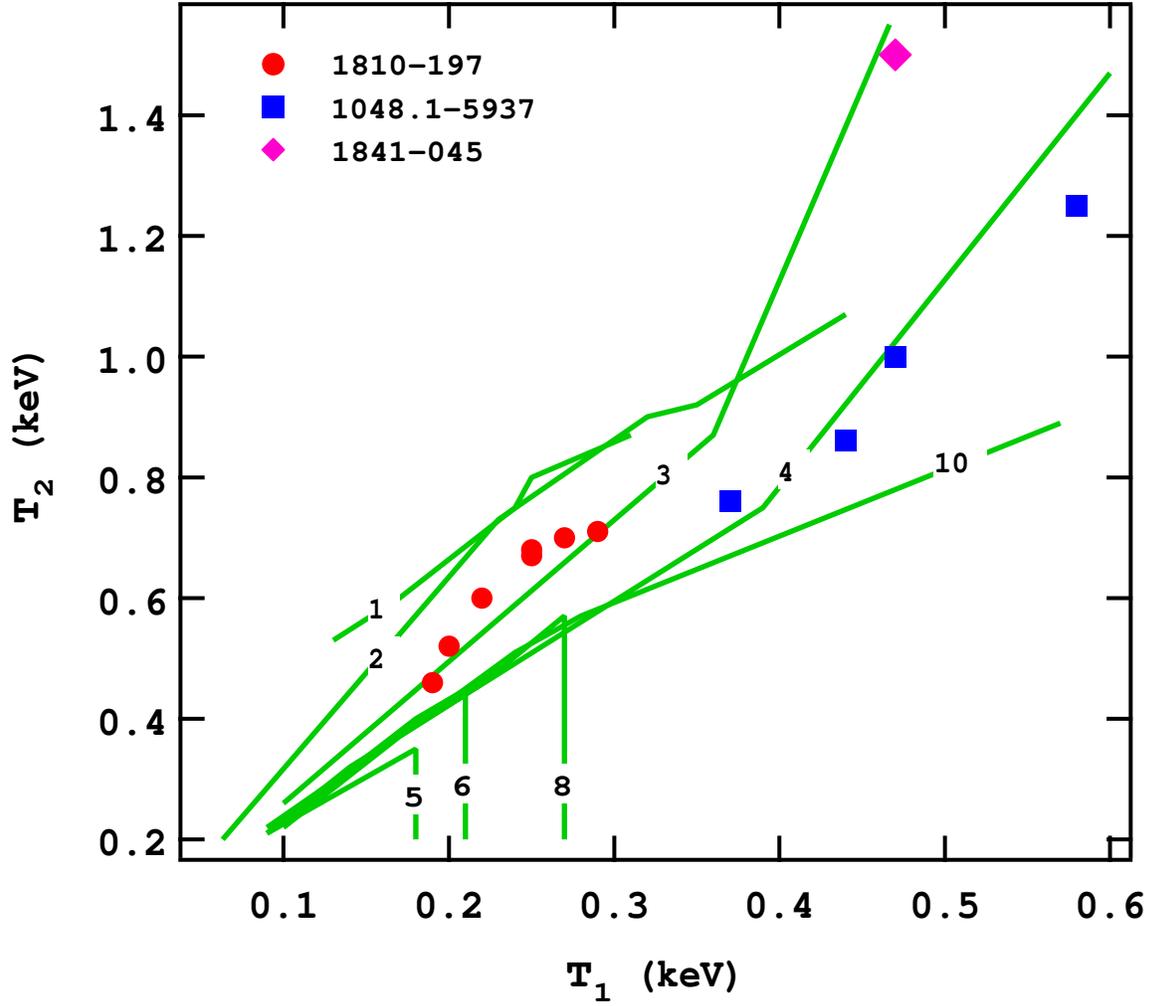}
   \caption{The temperatures of two black bodies used to fit our theoretical 
spectra for models with different magnetic field strengths in units of 
$10^{14}$~G (solid lines) as well as the temperatures inferred from 
multiple observations of three magnetars (filled markers). 
}

\label{obs}
\end{figure*}

The magnetic field leaves its imprint also on the continuum via both
atmospheric and magnetospheric processes. On the one hand, these two
processes work in tandem to strengthen the observable continuum
signature: Compton upscattering hardens the high energy excess already
present because of atmospheric effects. On the other hand, the exact
shape and hardness of the resulting spectrum depends also on the
strength of the unscattered cyclotron line, especially if it occurs at
energies above the peak of the X-ray emission. This is because, even
though the line is not visible as a distinct feature, it still causes
a suppression of the continuum spectrum. As a result, when we use a
phenomenological two-blackbody spectral model to describe the
theoretical spectra, the inferred blackbody temperatures depend
strongly and not monotonically on the magnetic field strength.  We
show this effect in Figure \ref{obs}. Note that we have not applied redshift
corrections to the temperatures obtained by fitting two blackbodies to
our theoretical spectra. Applying such a correction only shifts the
predicted correlations between the temperatures along the diagonal
without affecting the qualitative result. 

The qualitative dependence of the two blackbody temperatures $T_1$ and
$T_2$ changes dramatically at magnetic field strength $\sim 5 \times
10^{14}$~G, where the unscattered cyclotron line is the strongest (see
Table \ref{t_eq}). For all values of the magnetic field strength, the two
temperatures in the phenomenological model increase with increasing
effective temperature.  However, the strength of the second blackbody
initially decreases with increasing magnetic field, becoming
negligible at moderate effective temperatures for $B \sim 5\times
10^{14}$~G. At even stronger fields, the second blackbody again
becomes more prominent.

In Table \ref{t_con}, we show the temperatures of the two blackbodies that
represent the observed X-ray spectra of several AXPs and SGRs and
compare them in Figure~3 to the theoretical predictions. It is
remarkable that the temperatures of the two blackbodies necessary to
fit the source spectra lie in the same narrow range predicted by our
theoretical models. More importantly, for the two sources for which
multiple observations requiring different temperature pairs exist, the
data nearly follow constant field lines. This may provide insight into
the evolution of the magnetic field following strong bursts.

Fitting directly the actual theoretical models to the observed spectra
of magnetars will provide a good handle on their surface magnetic
field strengths. However, even the preliminary comparison shown in
Figure \ref{obs} results in magnetic field strengths that are
remarkably close to those inferred from the spin-down formula for
these sources: $2.7 \times 10^{14}$~G for XTE~1810$-$197
\citep{gh2006}, $(2.4-5) \times 10^{14}$~G for 1E~1048.1$-$5947
\citep{k2001}.  This implies that very weak (see Table~1) cyclotron
lines are likely to be present in the X-ray spectra of these sources.
Detection of such lines with future X-ray telescopes such as {\em
Constellation}-X will confirm these predictions with direct
measurement of the surface magnetic fields of magnetars.

\begin{deluxetable}{cccccc}
\tablecolumns{6}
\tablewidth{480pt}
\tablecaption{Observational upper limits on line equivalent widths}
\tablehead{Energy Range & Line Width & \multicolumn{4}{c}{Equivalent width (eV)} \\
(keV) 	     &    (eV)    & SGR 1806-20\tablenotemark{a} & SGR 1900+14\tablenotemark{a} & 
1E~1048.1-5947\tablenotemark{b} & 4U 0142+61\tablenotemark{c}}
\startdata
1 - 2        & 0        & -	&  15	& 10	& -	\\
             & 100      & -	&  50	& 20	& 15\\
             & 200      & -	&  50	& -	& - \\   
2 - 4        & 0        &  10   &  20	& 20	& -\\
  	     & 100	&  20	&  30	& 30	& - \\ 
             & 200	&  60	&  60	& -	& 70\\
4 - 5		&0	&  20	&  20	& 30	&\\
		& 100	&  30	&  35	& 55	&\\
		& 200	&  40	&  50	& -	&\\
		& 400	& -	& -	& -	& 200 \\
5 - 6		& 0	& 20	& 50	& 60	& - \\
		& 100	& 30	& 75	& 60	& - \\
		& 200	& 40	& 110	& -	& - \\
		& 500	& -	& -	& -	& 500 \\
6 - 7		& 0	&  10	&  90	& 90	&\\
		& 100	&  25	&  125	& 150	&\\
		& 200	&  30	&  170	& -	&\\
7- 8		& 0	&  10	&  35	& 150	&\\
		& 100	&  30	&  85	& 220	&\\
		& 200	&  30	&  100	& -	&\\
\enddata
\tablenotetext{a}{\citet{m2006}}
\tablenotetext{b}{\citet{t2006}}
\tablenotetext{c}{\citet{j2002}}
\label{uplim}
\end{deluxetable}

\begin{deluxetable}{cccc}
\tablecolumns{4}
\tablewidth{300pt}
\tablecaption{Inferred temperatures of two-blackbody fits to magnetar spectra}
\tablehead{Source Name 		& $kT_{1}$ & $kT_{2}$ & Reference\\
	    		& keV	& keV      &}
\startdata
1E1048.1-5947		& 0.47  & 1.00	& \citet{t2005} 			\\
			& 0.44	& 0.86	& 			\\
			& 0.37	& 0.76	& 			\\
			& 0.58	& 1.25	& \citet{o1998}		\\
1E1841-045    		& 0.47	& 1.50	& \citet{m2003}			\\
XTE 1810-197		& 0.25	& 0.68	& \citet{gh2006}		\\
			& 0.29	& 0.71	&					\\
			& 0.27	& 0.70	&					\\
			& 0.25	& 0.67	&					\\
			& 0.22	& 0.60	&					\\
			& 0.20	& 0.52	&					\\
			& 0.19	& 0.46	&					\\
\enddata
\label{bb_obs}
\end{deluxetable}

\acknowledgements T. G. wishes to thank to the members of the
University of Arizona Physics Department and Theoritical Astrophysics
Program for their hospitality during the visit to Tucson, Arizona.

\end{document}